\documentclass[aps,prd, twoside,superscriptaddress,showpacs, preprintnumbers, showpacs, nofootinbib, 
twocolumn]{revtex4-1}
\pdfoutput=1

\usepackage{amssymb,amsmath,latexsym,graphics, graphicx,epsfig,multirow,comment,hyperref,appendix, verbatim} 
\usepackage{pifont}
\usepackage{lipsum}
\usepackage{mathrsfs}  

\usepackage{slashed}
\usepackage{subfigure}
\usepackage{feynmp}
\usepackage{graphicx}
\usepackage{amsfonts}
\usepackage{color}
\usepackage{cancel}

\def\hbar{\hspace{0pt}\raisebox{1pt}{$-$} \hspace{-7pt} h}

\def\5{\overline 5}

\definecolor{JJ}{RGB}{0,144,255}

\newcommand{\be}{\begin{equation}}
\newcommand{\ee}{\end{equation}}
\newcommand{\bea}{\begin{eqnarray}}
\newcommand{\eea}{\end{eqnarray}}
\newcommand{\nn}{\nonumber}

\newcommand{\ba}{\begin{eqnarray}}
\newcommand{\ea}{\end{eqnarray}}

\begin{document}
\title{Unitarity implications of a diboson resonance in the TeV region for Higgs physics}

\author{Giacomo Cacciapaglia}

\affiliation{Universit\'e de Lyon, France; Universit\'e Lyon 1, Villeurbanne, France; CNRS/IN2P3, UMR5822, IPNL F-69622 Villeurbanne Cedex, France}
\email{g.cacciapaglia@ipnl.in2p3.fr}

\author{Mads T. Frandsen}

\affiliation{CP$^{3}$-Origins and the Danish Institute for Advanced Study, University of Southern Denmark, Campusvej 55, DK-5230 Odense M, Denmark}

\email{frandsen@cp3-origins.net}

\begin{abstract}
We investigate the implications of a putative new resonance in the TeV region coupled to the weak bosons. By studying perturbative unitarity of longitudinal $WW$ scattering, we find that a weakly coupled spin-1 resonance, that explains the ATLAS diboson excesses, is allowed with a SM-like Higgs. On the other hand, larger values of the resonance couplings, preferred in models of strong dynamics, would imply either sizeable reduction of the Higgs couplings or new physics, beyond the diboson resonance, at a few TeV.

\preprint{CP3-Origins-2015-026 DNRF90, DIAS-2015-26, LYCEN-2015-06}

\end{abstract}



\maketitle

\section{Introduction}
The recent ATLAS search for diboson resonances using boson-tagged jets \cite{Aad:2015owa} finds local excesses of 3.4, 2.6 and 2.9$\sigma$ in the $WZ$, $WW$ and $ZZ$ tagged boosted dijets with invariant mass spectrum around 2 TeV. 
The global significance is currently 2.5$\sigma$ and the required cross section at the border of null searches for diboson resonances in (semi-)leptonic decay modes as summarised in \cite{Franzosi:2015zra}, although see~\cite{Aguilar-Saavedra:2015rna}.
 
The general features, regarding resonance mass, cross-section and decay patterns, are broadly consistent with expectations from e.g. new electroweak scale spin-1 resonances \cite{Hisano:2015gna,Cheung:2015nha,Dobrescu:2015qna,Brehmer:2015cia,Gao:2015irw,Cao:2015lia}, including those from new strong dynamics at the electroweak scale \cite{Belyaev:2008yj,Fukano:2015hga,Franzosi:2015zra,Thamm:2015csa}.

In addition, CMS observes a number of local excesses near 2 TeV invariant masses at the $2\sigma$ level, e.g. in a boosted search for $WH$ with the Higgs decaying hadronically \cite{CMS:2015gla}.

The presence of a resonance in the TeV region coupled to the weak bosons of the Standard Model (SM) has implications for the perturbative unitarity of the theory. Here we illustrate this in the case of a spin-1 weak triplet resonance.
In particular the combination of dijet (and dilepton) limits and perturbative unitarity constrains the allowed couplings that can explain the excess without requiring modified Higgs couplings or additional new physics accessible at the LHC, or at a future 100 TeV collider.

\section{Simplified Lagrangian for dibosons from heavy spin-1 resonances}
We are interested in a new resonance $R$ coupled to the SM massive gauge bosons and to quarks - we will assume the production observed at ATLAS is via Drell-Yan. 
We will focus on the case that $R$ has spin-1, and postulate the presence of both charged and neutral components of $R$: under these assumptions, we can write a simple Lagrangian for its couplings to quarks (and leptons)

\bea
\label{Eq:fermcoupl}
\mathcal{L}^R_{\rm fermions}&=&\sum_{f_u,f_d} \bar{f}_u\slashed{\mathcal{R}}^+\left(g^V_{f+}- g^A_{f+}\, \gamma_5\right) f_d + {\rm h.c.} \nonumber
\\
&+&\sum_{f}  \bar{f} \slashed{\mathcal{R}}^0 \left(g^V_{f0}-g^A_{f0}\, \gamma_5\right) f \ ,
\eea
where $f=q,\ell$ runs over all SM fermions and $f_u$ ($f_d$) runs over all up-type (down-type) fermions. We have expressed both vertices in the vector-axial basis and suppressed flavour indices: in the following, for simplicity, we will focus on the case of universal couplings (that avoid any issue with flavour constraints).

The new vectors $R^0$ and $R^\pm$ also couple to the massive gauge bosons $W$ and $Z$:
in order to study the perturbative unitarity of the longitudinal $WW$ scattering, it is convenient to use the Equivalence Theorem and work in the electroweak unbroken phase directly with couplings to the Goldstone bosons of the electroweak sector. 
For simplicity, we assume that $R\equiv R^a$ is a triplet of weak isospin: a simple parameterisation of the couplings of $R$, and of the Higgs $h$, to the goldstone bosons $\phi$ is 
\begin{eqnarray}
{\cal L}_{R\phi\phi} & = & g_{R\phi\phi}\ \varepsilon^{abc}\ R^a_\mu\ \phi^b\ \partial^\mu\phi^c \ , \label{eq:LVpp_main} \\
{\cal L}_{h\phi\phi} & = & \frac{g_{h\phi\phi}}{F_\phi}\ h\ \partial^\mu\phi^a\ \partial_\mu\phi^a\,.
\end{eqnarray}
The second term corresponds to parameterising the couplings of the Higgs to gauge bosons in a non-linear formalism, and it describes possible deviations of the Higgs couplings from the SM values:
\begin{equation}
g_{h\phi\phi} = \frac{M_h}{F_\phi} \left(1 - \delta_{h} \right)\,,
\end{equation} 
where $M_h = 125$ GeV is the Higgs mass, $F_\phi$ can be identified with the electroweak scale $v=246$ GeV, and $\delta_{h} = 0$ corresponds to the SM limit.
A more complete study is given, e.g., in \cite{Foadi:2008xj}, but our conclusions will not depend on the above simplification.

The resulting isospin invariant $\phi-\phi$ scattering amplitude, including self-interactions, is given by
\begin{align}
A(s,t,u)&=\left(\frac{1}{F_\phi^2}
-\frac{3g_{R\phi\phi}^2}{M_R^2}\right)s
-\frac{g_{h\phi\phi}^2}{M_h^2}\frac{s^2}{s-M_h^2} \nonumber
\\
&-g_{R\phi\phi}^2\left(\frac{s-u}{t-M_R^2}+\frac{s-t}{u-M_R^2}\right) \ .
\label{eq:inv_2}
\end{align}
The scattering amplitude can be expanded in its isospin $I$ and spin $J$ components, $a^I_J$, with the $I=0$ and $J=0$ partial wave amplitude
\begin{multline}
a_0^0(s) = \frac{1}{64\pi} \int_{-1}^1 d\cos\theta \big[ 3A(s,t,u)+\\
A(t,s,u)+A(u,t,s)\big] \ ,
\end{multline}
 having the worst high energy behaviour.
The above amplitude depends on 4 parameters:
\begin{equation}
M_h , \quad \delta_{h} , \quad M_R , \quad g_{R\phi\phi} \nn
\end{equation}
and it should be bounded as $|a^0_0| < 1/2$ in order for the perturbative unitarity to be preserved.
From current Higgs searches we know the Higgs mass with an excellent precision, $M_h\simeq 125$ GeV, while the couplings to gauge bosons have been measured with an accuracy of a few tens of \% depending on the hypothesis on the couplings to fermions and eventual new light states.
We further set $M_R\sim 2$ TeV, in order to fit the ATLAS diboson excess.

In the following we will infer the allowed values of $g_{R\phi\phi}$ by imposing that the diboson production cross section matches with the excess observed by ATLAS: additional constraints can be imposed by the absence of an excess in other channels like dijet and dilepton searches. This procedure is rather general, and it can be repeated for any spin-1 state that couples to SM weak gauge bosons.
Requiring perturbative unitarity of the longitudinal $WW$ scattering can then be used to infer information about the couplings of the Higgs and on the presence of further light resonances.

In the model under consideration, the main production cross section is Drell-Yan from the resonance couplings to quarks: setting $g^V_{q+} = g^V_{q0} = 1$, and all other quark couplings to zero, the production cross section for the 2 TeV resonances at run-I is \cite{Franzosi:2015zra}
\begin{eqnarray}
\sigma_{\rm ref} &\equiv& \sigma (pp\to R^{\pm}) \simeq 1.7 \times 10^3 \ \textrm{fb} \\
 & \sim & \sigma (pp\to R^{0}) \simeq 1.5 \times 10^3 \ \textrm{fb}\,. \nn
\end{eqnarray}
Thus, the total cross-sections into diboson final states, in the narrow width approximation, are given by
\bea
\sigma_{W^+W^-} &=& \left( {g^V_{q0}}^2 +  {g^A_{q0}}^2\right)\ Br[R^0 \to W^+ W^-] \ \sigma_{\rm ref} \,, \\
\sigma_{W^+Z} &=& \left( {g^V_{q+}}^2 +  {g^A_{q+}}^2\right) \ Br[R^+ \to W^+ Z] \ \sigma_{\rm ref}\,.
\eea 
and the partial widths are given by
\begin{eqnarray}
\Gamma(R^{\pm/0}\rightarrow q\bar{q})&\simeq&\frac{N_f \left({g^V_{q+/0}}^2+{g^A_{q+/0}}^2\right)}{4\pi} 
M_R\,, \nn \\
\Gamma(R^{\pm/0}\rightarrow l\bar{l})&\simeq&\frac{N_f \left( {g^V_{l+/0}}^2+{g^A_{l+/0}}^2\right)}{12\pi} 
M_R\,, \\
\Gamma(R^{\pm/0} \rightarrow VV)&\simeq& \frac{g_{R\phi\phi}^2}{48\pi} M_R\,, \nn
\end{eqnarray}
in the limit of massless SM states, where the leptonic final states are $l\bar{l} = l^+ l^-$ and $\nu \bar{\nu}$ for the neutral state $R^0$ and $l^+ \nu$ for the charged $R^+$, and $N_f$ is the number of generations (assuming universal couplings). 
For the decays into gauge bosons, we again used the Equivalence Theorem: $\Gamma (R \rightarrow VV) = \Gamma (R \rightarrow \phi \phi)$ and finally we neglected decay modes into $HV$ although they are relevant in more complete models, e.g \cite{Zerwekh:2005wh,Belyaev:2008yj}.

Both ATLAS and CMS have searched for resonances in dijet and leptonic final states: such results can be used, together with the constraint $\sigma_{VV}\sim 10$ fb, to extract generic bounds on ratios of couplings.
To simplify the analysis, we will set to zero all axial couplings, and choose a single coupling for quarks, i.e. $g^V_{q0} = g^V_{q+} = g_q$, and a single coupling for leptons, $g^V_{l0} = g^V_{l+} = g_l$. Under this simplifying assumption, the partial widths of the charged and neutral vectors are the same, so that the two channels will give rise to approximately the same cross sections .
From run-I, we will use the following numerical constraints:
\begin{equation}
\sigma_{VV} \sim 10~\mbox{fb}\,, \quad \sigma_{qq} \lesssim 200~\mbox{fb}\,, \quad \sigma_{ll} \lesssim 0.5~\mbox{fb}\,.
\end{equation}
These numbers summarise the search limits from various studies of ATLAS and CMS, see e.g. \cite{Franzosi:2015zra}.
Ratios of these cross sections only depend, at leading order, on ratios of couplings, thus we can use them to extract direct constraints on the couplings (considering only the 2 light generations for dilepton limits and all quarks except the top for the dijet limits):
\begin{eqnarray}
\frac{\sigma_{VV}}{\sigma_{qq}} = \frac{g_{R\phi\phi}^2}{24 g_q^2} > \frac{1}{20} & \Rightarrow & g_{R\phi\phi} \gtrsim 1.2\, g_q\,, \\
\frac{\sigma_{ll}}{\sigma_{VV}} = \frac{8 g_{l}^2}{g_{R\phi\phi}^2} < 0.05 &\Rightarrow & g_{l} \lesssim 0.08\, g_{R\phi\phi}\,.
\end{eqnarray}

\begin{figure}[htb!] 
\begin{center}
 \includegraphics[width=.85\columnwidth]{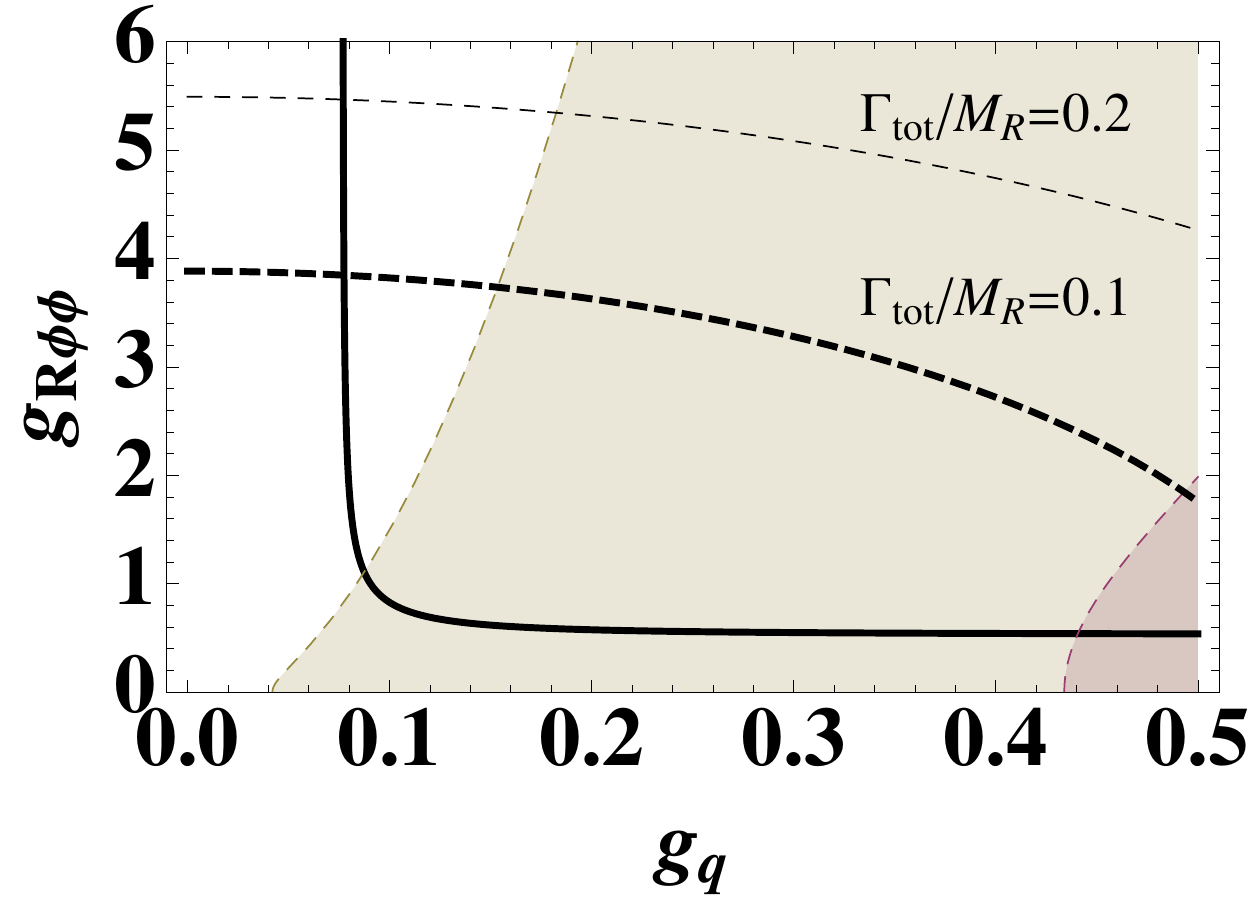}
\caption{Isoline $\sigma_{VV} = 10$ fb in the $g_q$--$g_{R\phi\phi}$ space (solid curve). The dark region in the lower right corner is excluded by $\sigma_{qq} < 200$ fb for $g_l = 0$. The lighter shaded region is excluded by $\sigma_{ll} < 0.5$ fb, assuming $g_q = g_l$. The dashed lines correspond to fixed values $\Gamma_{\rm tot}/M_R = 10$\% and $20$\%.}
\label{fig:bounds}
\end{center}
\end{figure}

We can now determine a lower limit on the value of $g_{R\phi\phi}$ by combining the requirement $\sigma_{VV}\sim 10$ fb and the constraints on the other channels. The result in the case where we set to zero the coupling to leptons is shown in Figure~\ref{fig:bounds}.
The figure shows that the coupling to quarks $g_q$ is only allowed to vary within a specific interval: the lower limit corresponds to a situation where $Br [R \to VV] = 100\%$, and thus $\sigma_{VV} = 10$ fb requires
\begin{equation}
g_{q} \geqslant \sqrt{\frac{10~\mbox{fb}}{\sigma_{\rm ref}}}\sim 0.077 \,.
\end{equation}
The upper limit on $g_q$, and lower limit on $g_{R\phi\phi}$, derives from saturating the bound from dijet searches, and leads to
\begin{equation}
g_q \lesssim 0.38\,, \quad g_{R\phi\phi} \gtrsim 0.47\,.
\end{equation}
If we assume that the coupling to quarks and leptons are of the same order, as it may happen schematically if the couplings originate from mixing of $R$ with the SM gauge bosons, then the constraints on dilepton resonances would strengthen the lower bound on $g_{R\phi\phi}$ and the upper bound on $g_q = g_l$:
\begin{equation}
g_q \lesssim 0.087\,, \quad g_{R\phi\phi} \gtrsim 1.1\,.
\end{equation}
It should be noted at this point that for $g_q$ close to the minimal value, and thus for large $g_{R\phi\phi}$, the $\sigma_{VV}$ depends very mildly on the precise value of the coupling to goldstone bosons, therefore fitting the excess cross section does not at present allow to extract precise information on $g_{R\phi\phi}$. 
This is particularly true for models of strong dynamics, where large $g_{R\phi\phi}$ are expected. The only constraints may therefore come from the indirect dependence of $g_q$ on $g_{R\phi\phi}$, which is however model dependent. As an example we can use the simplest resonance extension of the chiral Lagrangrian, based on the ``hidden local symmetry'' formalism \cite{Bando:1984ej,Bando:1987br} applied to heavy spin-1 vectors and dynamical symmetry breaking in \cite{Casalbuoni:1985kq}. Here the charged vector couples to the left-handed quarks with strength
\bea
g^V_{q+} = g^A_{q+} \sim \frac{g^2}{4 \sqrt{2} g_\rho} \simeq \frac{0.09}{g_\rho}
\eea
leading to $g_\rho \sim 1.4$ from fitting the excess. The relation of $g_\rho$ with $g_{R\phi\phi}$ then gives 
\bea
g_{R\phi\phi} = \frac{M_R^2}{2 g_\rho v^2}  \simeq 23\,,
\eea
thus showing that the required coupling is too large for the model to make sense. However, in less minimal models, the relation between the couplings can be relaxed, leading to different predictions, e.g \cite{Zerwekh:2005wh,Belyaev:2008yj}. 

Here we focus on the constraints on $g_{R\phi\phi}$ arising from perturbative unitarity of $WW$ scattering.
Another quantity sensitive to large values of $g_{R\phi\phi}$ is the total width of the resonance.
In Figure \ref{fig:bounds} we show contours of fixed $\Gamma_{\rm tot}/M_R$: this parameter is important to check that the narrow width approximation, that has been used to derive the constraints, is still valid. 

The results show that a large coupling of the new resonance to $W$ and $Z$ is needed, especially when the resonance is not leptophobic: it is then relevant to consider the production cross section via vector boson fusion (VBF), which may enhance the signal. We reinterpreted the results from \cite{Falkowski:2011ua} to obtain
\begin{equation}
\sigma_{VBF} = g_{R\phi \phi}^2 \times 0.1~\mbox{fb}\,.
\end{equation}
This cross section is rather small, and it will only contribute significantly to the excess for very large values of the coupling $g_{R\phi \phi}$. On the other hand, if run-II of the LHC confirms the excess, adding a VBF jet-tag may allow to measure the VBF component versus the Drell-Yan one, and give direct access to the value of the $g_{R\phi\phi}$ coupling.

Armed with the above results, we can now study the perturbative unitarity in the longitudinal $WW$ scattering, which only depends on $\delta_h$ and $g_{R\phi\phi}$, once the mass of the resonance is fixed at $M_R = 2$ TeV.
Requiring perturbative unitarity of $a_0^0(s)$, we are then able to determine, for each values of $g_{R\phi\phi}$, a relation between the deviation of the Higgs couplings and $\Lambda_{NP}$, the scale where perturbative unitarity is lost - and hence where additional new physics is required.
Expanding the amplitude for large $s$, the leading order result is 
\begin{align}
a_0^0(s)&=\frac{1}{16\pi } (\frac{1}{F_\phi^2}- \frac{g_{h\phi\phi}^2}{M_H^2} - \frac{3g_{R\phi\phi}^2  F_\phi^2}{M_R^2} )s + ...
\\
&\simeq \frac{1}{16\pi F_\phi^2}  (2 \delta_{h} - \delta_h^2 - 3g_{R\phi\phi}^2  \frac{ F_\phi^2}{m_R^2} ) s  + ...
\end{align}
where $...$ contains the constant piece, and pieces falling with energy. The term linear in $s$ is cancelled for
\begin{equation}
\delta_h \simeq 0.02\, g_{R\phi\phi}^2\,,
\end{equation}
thus one can only hope to tame the growing amplitude if the coupling of the Higgs is reduced with respect to the SM value, as expected. 

\begin{figure}[htb!] 
\begin{center}
  \includegraphics[width=.85\columnwidth]{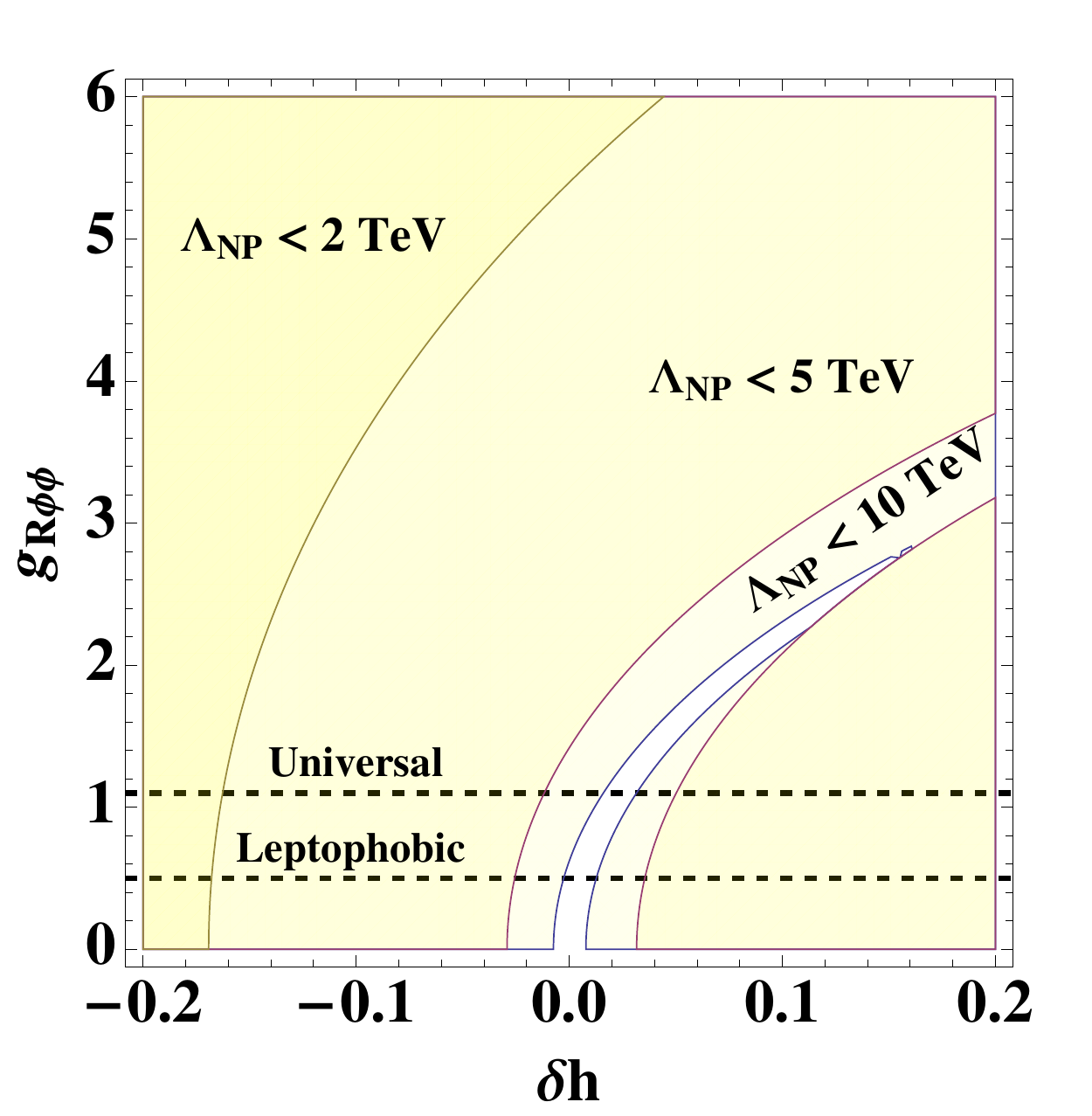}
\caption{Contours of fixed scale where perturbative unitarity is lost, $\Lambda_{\rm NP} = 2$, $5$ and $10$ TeV, in the plane $\delta_h$ vs $g_{R\phi\phi}$. The region in white is where the cut-off is above $10$ TeV. The horizontal lines mark the lower bounds on $g_{R\phi\phi}$ for a leptophobic $R$, and in the case of equal coupling to leptons and quarks.}
\label{fig:Uni1}
\end{center}
\end{figure}

In Fig.~\ref{fig:Uni1} we plot contours of equal $\Lambda_{\rm NP}$, the scale where new states are expected, in the plane $\delta_h$--$g_{R\phi\phi}$. 
We see that for low values of $g_{R\phi\phi}$, allowed when the resonance is leptophobic, the Higgs can still be very SM-like and the cut-off above $10$ TeV. For values between $1 \lesssim g_{R\phi\phi} \lesssim 3$, there can still be no additional new states up to 10 TeV, however at the price of a modification of the Higgs coupling to dibosons at the level of 10\%. For larger values of $g_{R\phi\phi}$, the cut-off from perturbative unitarity cannot be above 10 TeV, thus one should expect new physics to show up at such energies possibly accessible during run-II. It is also interesting to notice the complementarity between the measurement of the Higgs couplings and information obtainable from perturbative unitarity: for instance, if the Higgs couplings were constrained within 10\% of the SM value, we would be able to infer that for $g_{R\phi\phi} \gtrsim 6$ some other states at or below the mass of $R$ are present.

\section{Conclusions}

The recent excesses in boosted diboson final states found by ATLAS in run-I data have stirred interest, because the indicated $\sim 2$ TeV mass scale and the production cross-sections agree with those obtained in models with additional spin-1 resonances coupled to the electroweak sector. Even though the excesses may be a statistical fluctuation, it is instructive to entertain the idea that it is real. Constraints on the production cross sections and couplings have been widely studied, thus in this work we focus on the implications for perturbative unitarity of longitudinal $WW$ scattering. 

In fact, the production cross-section of the resonance is quadratically sensitive to its  quark couplings (assuming Drell-Yan production) and these couplings are directly constrained by dijet searches. In most models of course the dilepton searches will provide an even better constraint. 

On the other hand the coupling of the resonance to dibosons is sensitively constrained by the $WW$ scattering amplitude 
independently of the resonance coupling to quarks. We thus used the contributions of the Higgs boson and the new resonance to this amplitude, to derive a constraint on the allowed deviation of the Higgs couplings to $WW$ and $ZZ$ as a function of the value of the resonance coupling to dibosons. We did this by requiring perturbative unitarity up to a certain scale. 
We find that for the smallest values of the resonance coupling to dibosons that explain the ATLAS data, roughly $0.5$, a very SM-like Higgs and no new physics below 10 TeV are allowed.
On the other hand, in the region of large coupling, where the resonance can still be narrow, e.g. a value of $\gtrsim 3$, the deviations in the Higgs couplings must either be in the 10\% range or additional new physics can be expected at the LHC.
For intermediate values of the coupling we find deviations in the Higgs coupling at the $\%$-level and up to a 10\% if we assume no additional new physics is present up to the 10 TeV scale.

While we focused our analysis on the recent 2 TeV diboson excesses, and their interpretation in terms of a spin-1 resonance, our conclusions are general and can be applied to any signal of new physics appearing in diboson final states at run-II of the LHC.

\bibliographystyle{JHEP}

\bibliography{wz.bib}

\end{document}